\documentclass[final,5p,times,preprint]{elsarticle}

\usepackage{amsfonts}
\usepackage{amsmath}
\usepackage{amssymb}
\usepackage{textgreek}
\usepackage{bm}
\usepackage{comment}
\usepackage{xcolor}
\usepackage{booktabs}
\usepackage[normalem]{ulem}
\usepackage[export]{adjustbox}
\graphicspath{{./figures/}}
\journal{Physics Letters B}

\usepackage[utf8]{inputenc}
\usepackage[english]{babel}
\biboptions{comma,sort&compress}

\usepackage{amsmath}
\usepackage{graphicx}
\usepackage[colorinlistoftodos]{todonotes}
\usepackage[colorlinks=true, allcolors=blue]{hyperref}
\usepackage[normalem]{ulem} 

\begin{document}

\begin{frontmatter}

\title{Searching for universality of dineutron correlation at the surface of Borromean nuclei}

\author[cea]{A.~Corsi}
\cortext[mail]{Corresponding author}
\ead{acorsi@cea.fr}
\author[rik,cns,tuda]{Y.~Kubota}
\author[FAMN]{J. Casal}
\author[FAMN]{M.~Gómez-Ramos}
\author[FAMN]{A.~M.~ Moro}
\author[cea]{G. Authelet}
\author[rik]{H. Baba}
\author[tuda]{C. Caesar}
\author[cea]{D. Calvet}
\author[cea]{A. Delbart}
\author[cns]{M. Dozono}
\author[key]{J. Feng}
\author[ipno]{F. Flavigny}
\author[cea]{J.-M. Gheller}
\author[lpc]{J. Gibelin}
\author[cea]{A. Giganon}
\author[cea]{A. Gillibert}
\author[toh]{K. Hasegawa}
\author[rik]{T. Isobe}
\author[miy]{Y. Kanaya}
\author[miy]{S. Kawakami}
\author[ehw]{D. Kim}
\author[cns]{Y. Kiyokawa}
\author[cns]{M. Kobayashi}
\author[tod]{N. Kobayashi}
\author[toh]{T. Kobayashi}
\author[tit]{Y. Kondo}
\author[dae,rik,atom]{Z. Korkulu}
\author[tod]{S. Koyama}
\author[cea]{V. Lapoux}
\author[miy]{Y. Maeda}
\author[lpc]{F. M. Marqu\'es}
\author[rik]{T. Motobayashi}
\author[tod]{T. Miyazaki}
\author[tit]{T. Nakamura}
\author[kyo]{N. Nakatsuka}
\author[kyu]{Y. Nishio}
\author[cea,tuda]{A. Obertelli}
\author[kyu]{A. Ohkura}
\author[lpc]{N. A. Orr}
\author[cns]{S. Ota}
\author[rik]{H. Otsu}
\author[tit]{T. Ozaki}
\author[rik]{V. Panin}
\author[tuda]{S. Paschalis}
\author[cea]{E. C. Pollacco}
\author[tum]{S. Reichert}
\author[cea]{J.-Y. Rousse}
\author[tit]{A. T. Saito}
\author[kyu]{S. Sakaguchi}
\author[rik]{M. Sako}
\author[cea]{C. Santamaria}
\author[rik]{M. Sasano}
\author[rik]{H. Sato}
\author[tit]{M. Shikata}
\author[rik]{Y. Shimizu}
\author[kyu]{Y. Shindo}
\author[dae,rik,atom]{L. Stuhl}
\author[rik]{T. Sumikama}
\author[cea,tuda]{Y.L. Sun}
\author[kyu]{M. Tabata}
\author[tit]{Y. Togano}
\author[tit]{J. Tsubota}
\author[rik]{T. Uesaka}
\author[rik]{Z. H. Yang}
\author[kyu]{J. Yasuda}
\author[rik]{K. Yoneda}
\author[rik,key]{J. Zenihiro}

\address[cea]{D\'epartement de Physique Nucl\'eaire, IRFU, CEA, Universit\'e Paris-Saclay, F-91191 Gif-sur-Yvette, France}

\address[rik]{RIKEN Nishina Center, Hirosawa 2-1, Wako, Saitama 351-0198, Japan}

\address[cns]{Center for Nuclear Study, University of Tokyo, Hongo 7-3-1, Bunkyo, Tokyo 113-0033, Japan}

\address[ECT]{European Centre for Theoretical Studies in Nuclear Physics and Related Areas (ECT$^*$), Villa Tambosi, Strada delle Tabarelle 286, I-38123 Trento, Italy}
\address[pd]{Dipartimento di Fisica e Astronomia ``G.~Galilei'' and INFN - Sezione di Padova, Via Marzolo 8, I-35131 Padova, Italy}
\address[FAMN]{Departamento de F\'{\i}sica At\'omica, Molecular y Nuclear, Facultad de F\'{\i}sica, Universidad de Sevilla, Apartado 1065, E-41080 Sevilla, Spain}

\address[tuda]{Department of Physics, Technische Universitat Darmstadt}
\address[pek]{Department of Physics, Peking University}
\address[ipno]{Institut de Physique Nucleaire Orsay, IN2P3-CNRS, F-91406 Orsay Cedex, France}
\address[lpc]{LPC Caen, ENSICAEN, Universite de Caen, CNRS/IN2P3, F-14050 Caen, France}
\address[toh]{Department of Physics, Tohoku University, Aramaki Aza-Aoba 6-3, Aoba, Sendai, Miyagi 980-8578, Japan}
\address[miy]{Department of Applied Physics, University of Miyazaki, Gakuen-Kibanadai-Nishi 1-1, Miyazaki 889-2192, Japan}
\address[ehw]{Department of Physics, Ehwa Womans University}
\address[tod]{Department of Physics, University of Tokyo, Hongo 7-3-1, Bunkyo, Tokyo 113-0033, Japan}
\address[tit]{Department of Physics, Tokyo Institute of Technology, 2-12-1 O-Okayama, Meguro, Tokyo 152-8551, Japan}
\address[atom]{MTA Atomki, P.O. Box 51, Debrecen H-4001, Hungary}
\address[kyo]{Department of Physics, Kyoto University, Kitashirakawa, Sakyo, Kyoto 606-8502, Japan}
\address[kyu]{Department of Physics, Kyushu University, Nishi, Fukuoka 819-0395, Japan}
\address[osa]{Research Center for Nuclear Physics, Osaka University, 10-1 Mihogaoka, Ibaraki, Osaka 567-0047, Japan}
\address[tum]{Department of Physics, Technische Universitat Munchen}
\address[dae]{Center for Exotic Nuclear Studies, Institute for Basic Science, Daejeon 34126, Republic of Korea}
\address[key]{School of Physics and State Key Laboratory of Nuclear Physics and Technology, Peking University, Beijing 100871, China}


\begin{abstract}
The dineutron correlation is systematically studied in three different Borromean nuclei near the neutron dripline, $^{11}$Li, $^{14}$Be and $^{17}$B, via the $(p,pn)$ knockout reaction measured at the RIBF facility in RIKEN. For the three nuclei, the correlation angle between the valence neutrons is found to be largest in the same range of intrinsic momenta, which can be associated to the nuclear surface. This result reinforces the prediction that the formation of the dineutron is universal in environments with low neutron density, such as the surface of neutron-rich Borromean nuclei.
\end{abstract}

\begin{keyword}
quasi-free scattering \sep Borromean nuclei \sep three-body model \sep Jacobi coordinates \sep dineutron
\end{keyword}

\end{frontmatter}

\section{Introduction \label{sec:intro} }
Halo nuclei appear close to the neutron dripline and present a diffuse matter distribution due to the reduced binding energy of the valence neutrons \cite{han87,tan13}. Nuclei formed by a core and two loosely bound neutrons, such that the subsystem formed by the core and the neutron is unbound, are called Borromean. Most of them present nuclear halos near the neutron dripline. 
Some examples are $^6$He, $^{11}$Li, $^{14}$Be and $^{17,19}$B. The correlation between the 
neutrons plays an essential role to stabilize these nuclei and has been the subject of a number of studies \cite{hagino05,hag15, cat84}.  We discuss here a specific form of spatially localized pairing correlation called dineutron \cite{mig73}. 
The strength of the pairing correlation evolves with density, going from the BCS (Bardeen–Cooper–Schrieffer) regime of loosely spaced correlations to the BEC (Bose-Einstein Condensate) regime of compact space correlation with decreasing density. 
This regime is expected to appear at the surface of neutron-rich nuclei, where neutron density is 10$^{-4}$ to 0.5 of the saturation density \cite{mat06,hag07}. 
Its onset appears to be strongly linked to the admixture of different parities in the wavefunction describing the valence neutrons \cite{cat84}. Halo nuclei, with their diffuse matter distribution, are an ideal probe to study this low-density correlation. The dineutron was 
experimentally revealed in $^6$He \cite{oga99,sun21} and $^{11}$Li \cite{nak06,sim99,sim07,kub20}, however the experimental evidence is still scarce.
Typically, dineutron correlation is explored via the opening angle between the two neutrons \cite{nak06,ber07}. 
An opening angle below 90$^\circ$ (90$^\circ$ corresponding to the non-correlated case) in coordinate space or above 90$^\circ$ in momentum space \cite{hag14,hag16b} points to a strong spatial correlation yielding a compact configuration. Intuitively, the opening angle in coordinate space is related to the inter-nucleon distance. References \cite{nak06} measured the $E1$ strength after Coulomb dissociation of $^{11}$Li, and extracted the opening angle in coordinate space based on the cluster sum rule and assuming an inert core \cite{esb92}. For $^{11}$Li the average angle obtained was $\theta$ = 48$^{+14}_{-18}$ $^\circ$, corresponding to a strong dineutron correlation. A more refined estimation can be obtained if the average neutron-neutron distance is measured independently. In such a way the authors of Ref. \cite{Hag07b} deduced from the $B(E1)$ measurement of Ref. \cite{nak06} a value of the mean opening angle of $\sim$ 56.2$^\circ$.
The average neutron-neutron separation was estimated via measurements of the two-neutron correlation function in dissociation reactions. This method has been applied to $^6$He, $^{11}$Li, $^{14}$Be \cite{mar00,mar01}. A combined analysis of the $B(E1)$ measurement \cite{nak06} and the correlation function \cite{mar01} has given a somewhat larger value of the opening angle in $^{11}$Li, $\sim 66^\circ$ \cite{ber07}, corresponding to a reduced dineutron correlation. Two-neutron transfer reactions can also be used to study $nn$ correlations~\cite{hag15}. In Ref.~\cite{tan08}, it was shown that the description of $^{11}$Li data required a model with a large pairing correlation.

Nucleon removal reactions are another method to access the opening angle \cite{chu97,sim99,mar01,sim07}. The authors of Ref. \cite{sim07} measured an opening angle in momentum space of $103.4\pm2.1^\circ$ for $^{11}$Li, and suggest that a dineutron configuration exists also in $^{14}$Be, 
although less developed than in $^{11}$Li. In both nuclei, the dineutron appears due to the mixing of different-parity orbitals \cite{cat84,sim99}. In contrast, the structure of the halo in $^{17}$B is mainly of $d$-wave character with small $s$-wave admixture \cite{yan21}, which should hinder the development of the dineutron correlation as both orbitals have the same parity. We note that the study of these correlations via breakup and knockout reactions is complementary to that performed through 2$n$ decay~\cite{Hagino16,Wang21,Spyrou12}, which has been used to explore the properties of two-neutron unbound systems.

Recently, Ref. \cite{kub20} 
introduced a new method based on quasi-free scattering reactions to study dineutron correlation as a function of its peripherality, i.e., distance from the baricenter of the system, and applied it to $^{11}$Li. The observable related to the peripherality is the intrinsic momentum of the removed nucleon in quasi-free scattering reactions. Being a fast removal process, the impact of final-state interactions on the observable of interest is assumed to be reduced, making its interpretation more straight-forward.
In this work we search for dineutron correlations applying this same method to $^{14}$Be and $^{17}$B, measured in the same experiment as Ref. \cite{kub20}. Our goal is to assess whether such correlation appears as a general feature at the surface of neutron-rich Borromean nuclei. The data is compared to calculations using a three-body model for the projectile and a quasi-free sudden model to describe the knockout process \cite{cas21}.

\section{\label{sec:level2} Experimental results}
\subsection{Setup}
The experiment was performed at the Radioactive Isotope Beam Factory operated by the RIKEN Nishina Center and the Center for Nuclear Study (CNS) of the University of Tokyo. Secondary
beams were produced using projectile fragmentation of a $^{48}$Ca primary beam at 345 MeV/nucleon with a typical intensity of 400 particle nA on a Be target. Fragmentation products were separated, detected and identified via the BigRIPS fragment separator \cite{fuk13}.
The cocktail beam was composed by 
$^{11}$Li, $^{14}$Be, and $^{17}$B, with a percentage of $\sim$ 80\%, 12\%, and 8\%, respectively. It 
impinged on the secondary target with an average energy of 246, 265 and 277 MeV/nucleon, respectively. The secondary target was the 15-cm thick liquid hydrogen target from the MINOS device \cite{obe14}, and was surrounded by a Time Projection Chamber acting as vertex tracker 
together with the beam tracking MWDC detectors. The detection system included the WINDS array of plastic scintillators for knockout neutron detection, and a MWDC followed by an array of plastic scintillators for the recoil proton detection. Those two detectors were key for the measurement of the intrinsic momentum of the removed neutron and the opening angle in the $(p,pn)$ reaction. The standard SAMURAI setup consisting of a set of drift chambers, the SAMURAI dipole magnet and two hodoscope walls was used for fragment analysis \cite{kob13}. The neutrons emitted at forward angles were detected by the NEBULA plastic scintillator array \cite{nak16}. 
We evaluated the acceptance cut they induce on the measurement of the intrinsic momentum and opening angle distribution using a Geant4 simulation. No bias is introduced by the experimental setup on the opening angle distribution, while the acceptance decreases for increasing intrinsic momentum (leading to off-plane scattering), as shown in the inset of Fig. \ref{f:theo-kmiss-dist}. 
More details on the rest of the setup can be found in Ref. \cite{cor19, kub20} and references therein.

\subsection{\label{sec:level3} Dineutron correlation}
\begin{figure}[t]
\begin{center}
     \begin{tikzpicture}
     \node at (0,0) {\includegraphics[width=0.9\linewidth]{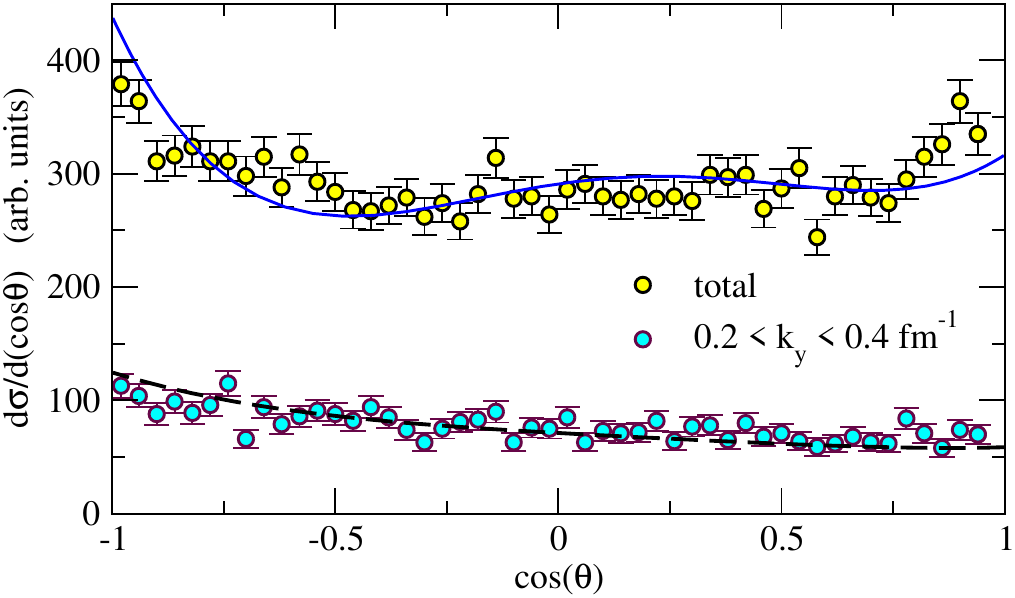}};
     \node at (-2.2,0.1) {\includegraphics[width=0.150\linewidth]{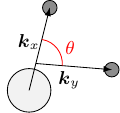}};
     \end{tikzpicture}
     
    \caption{\label{f:theo-angle-dist} Correlation angle distribution for $^{14}$Be (integrated in $E_\mathrm{rel}$), compared with our theoretical calculation. The lower points and dashed line correspond to an intrinsic momentum window of $0.2 < k_y < 0.4$ fm$^{-1}$. A schematic representation of the angle in Jacobi coordinates is embedded in the figure.}
 \end{center}
\end{figure}

The measurement of the momenta of the outgoing proton and removed neutron allows to reconstruct the intrinsic momentum of the neutron before removal (within the quasi-free approximation):\\
\begin{equation}
    \vec{k_y} := \vec{k_{n1}} = \vec{k_{n1}'} + \vec{k_p'} - \vec{k_p}  \\   
\end{equation}

where $k_{n1}$ ($k_{n1}^{'}$) is the momentum of the neutron in the initial (final) state and $k_p$ ($k_p^{'}$) the one of the target (recoil) proton.
The correlation angle, or the opening angle $\theta$ in momentum space, is the angle between the Jacobi momenta $k_x$ and $k_y$: \\
\begin{equation}
    cos(\theta)=\frac{\vec{k_x} \cdot \vec{k_y}}{|\vec{k_x}||\vec{k_y}|}
\end{equation}
with 
\begin{equation}
    \vec{k_x}=\vec{k'_{n2}}-\vec{k'_{f}}
\end{equation}
where $k'_{n2}$, $k'_{f}$ are the momenta of the remaining valence neutron and fragment in the final state. This representation of the three-body system in terms of Y Jacobi coordinates is illustrated in the inset of Fig. \ref{f:theo-angle-dist}. 
In the following, we illustrate the intrinsic momentum and correlation angle distribution for the case of $^{14}$Be. Both are compared with a theoretical calculation performed within a quasi-free sudden model \cite{cas21,kik16}, using the three-body model for $^{14}$Be from \cite{cor19}. 

\begin{figure}[t]
\begin{center}

     \includegraphics[width=1\linewidth]{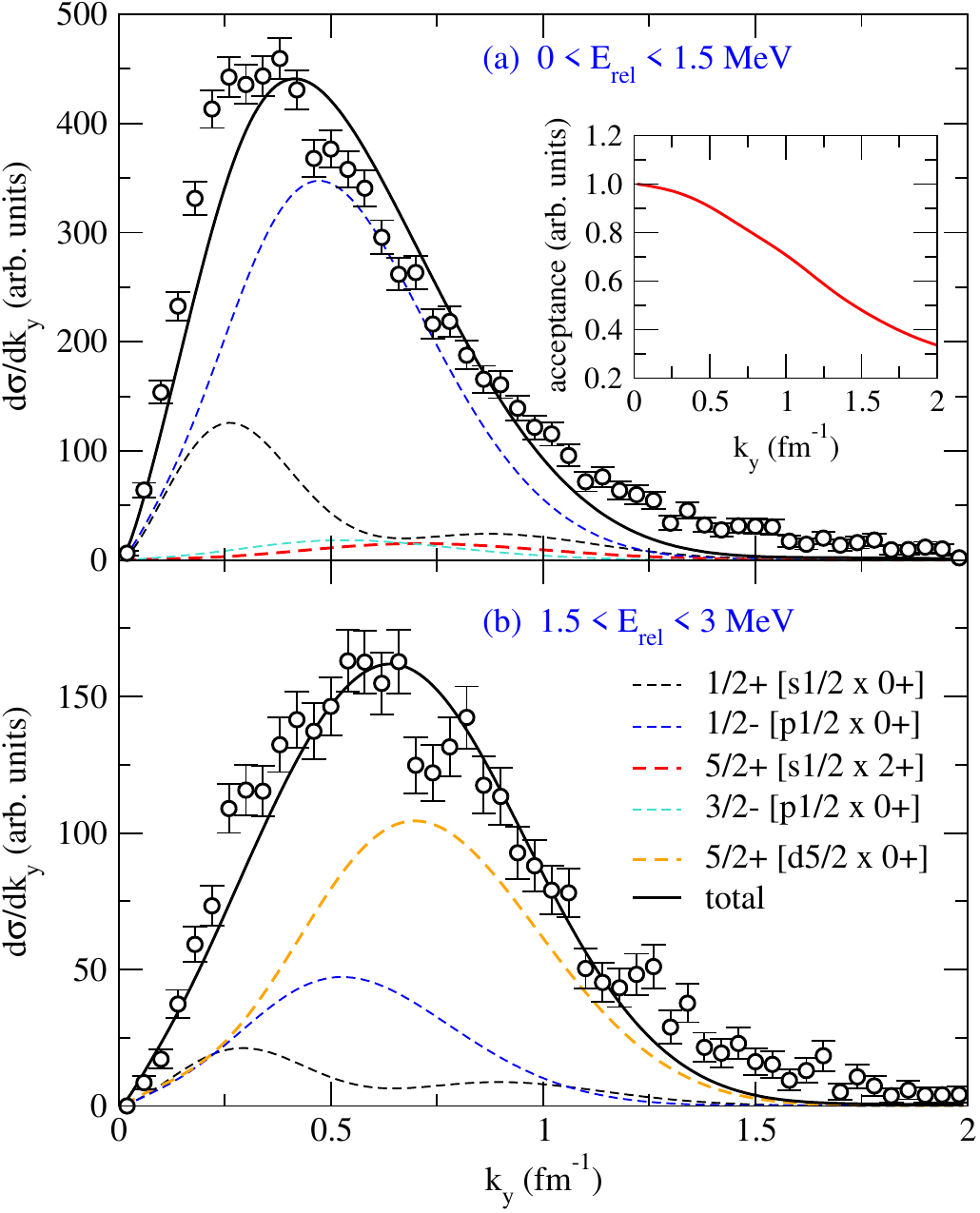}
    \caption{\label{f:theo-kmiss-dist} Comparison between experimental and theoretical intrinsic momentum distribution for the knocked-out neutron of $^{14}$Be for (a) $0<E_{\rm rel}<1.5$ MeV and (b) $1.5<E_{\rm rel}<3.0$ MeV in the $^{13}$Be system. The dashed lines correspond to the contribution from different $^{13}$Be states (see the text for details), and the solid line is the total. The inset shows the acceptance. }
 \end{center}
\end{figure}

Figure~\ref{f:theo-kmiss-dist} shows the intrinsic momentum distribution of the removed nucleon for two different relative-energy intervals in the $^{13}$Be system. The theoretical distributions are already corrected for the experimental acceptance (see inset of Fig.~\ref{f:theo-kmiss-dist}a) and convoluted with the experimental resolution of 0.17 fm$^{-1}$. Each relative-energy ({$^{12}\mathrm{Be}+n$}) interval encompasses a peak in the spectrum of $^{13}$Be \cite{cor19}. 
The comparison to theoretical calculations show that the 0-1.5 MeV interval is dominated by the $p_{1/2}$ component (72\% of the total in this energy range), while the 1.5-3 MeV interval is dominated by the $d_{5/2}$ component (60\% of the total). 
This is consistent with the interpretation of the relative-energy spectrum of $^{13}$Be provided in Ref. \cite{kon10, cor19} as composed of a $p$-wave resonance centered at 0.5 MeV followed by a broader $d$-wave resonance. The different lines in Fig.~\ref{f:theo-kmiss-dist} are labeled as $J^\pi[\ell_j \otimes S_c]$, where the single-particle angular momentum $\ell_j$ couples with the spin of the core $S_c$ to give the total angular momentum $J^\pi$ of the binary subsystem $^{13}$Be after knockout. Note that, since the ground state of $^{14}$Be is a 0$^+$ state, the angular momentum of the knocked-out neutron has to match $J^\pi$, e.g., $5/2^+$ contributions correspond to a $d$-wave knockout. It is worth noting that the calculations presented in Fig.~\ref{f:theo-kmiss-dist} are not a fit to the experimental data but the results of the structure model (and corresponding partial-wave content) of Ref.~\cite{cor19}, therefore the agreement is not perfect. In particular the disagreement in the peak in the lower energy range may suggest a larger $s$-wave component. However, in \cite{cor19}, an increase in $s$-wave led to a worse description of the low-energy distribution. 
Similarly, there is a slight disagreement for the largest $k_y$ values that may be associated to missing components in the wave function due to limitations of the model, as discussed in \cite{cor19} for large relative energies.

Fig.~\ref{f:theo-angle-dist} shows the correlation angle distribution integrated over all intrinsic momenta, and for intrinsic momenta between 0.2 fm$^{-1}$ and 0.4 fm$^{-1}$. One can see that the inclusive distribution is rather symmetric, while an asymmetry appears for some range of values of the intrinsic momentum. The calculations are able to capture this behaviour. The range between 0.2 fm$^{-1}$ and 0.4 fm$^{-1}$ is the one yielding the maximum asymmetry with an enhancement of values of the correlation angle larger than 90$^\circ$.  
This points towards a geometrically compact configuration of the two-neutron system (the dineutron) at low intrinsic momenta, which can be associated to the nuclear surface. 

\begin{figure}[h!]
\begin{center}
     \includegraphics[width=0.5\textwidth]{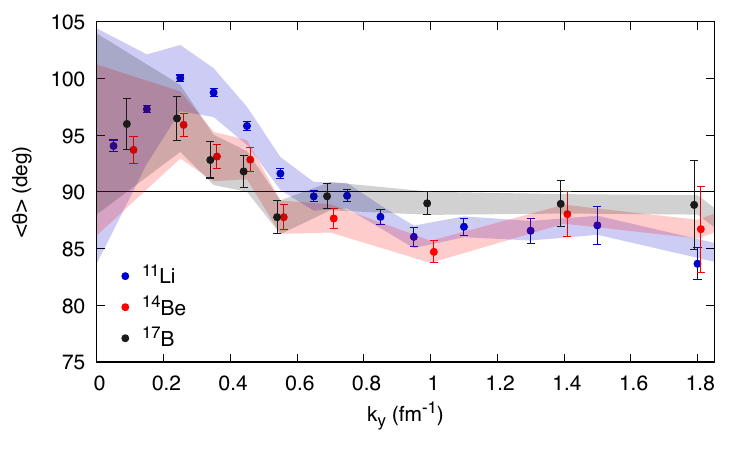}
    \caption{\label{f:angle} Average correlation angle as a function the intrinsic momentum $\vec{k}_y$ for $(p,pn)$ reaction on $^{11}$Li, $^{14}$Be and $^{17}$B. The points are presented with their statistical error (bars) and systematic error (band). Points for $^{14}$Be and $^{17}$B are slighty offsetted along the x-axis for clarity.}
\end{center}
\end{figure}

The average correlation angle $\theta$, obtained taking event by event the arccos of the data plotted in Fig. \ref{f:theo-angle-dist}, is plotted as a function of the intrinsic momentum ranging from 0 to 1.8 fm$^{-1}$ in Fig.~\ref{f:angle}. The data for $^{14}$Be are compared to the ones for $^{11}$Li and $^{17}$B measured in the same experiment \cite{kub20, cor19, yan21}. 
The nucleus of $^{11}$Li is considered as a reference case of well developed dineutron correlation \cite{nak06,sim07,kub20} so, as expected, it presents the largest deviation from 90$^\circ$. 
It is however remarkable that for both $^{14}$Be and $^{17}$B the data also show a significant deviation in the correlation angle distribution in the same range of momenta. This deviation points to the appearance of a dineutron correlation for intrinsic momenta smaller than 0.4 fm$^{-1}$, which corresponds to the nuclear surface \cite{kub20}, for all measured nuclei. We note that the larger value of the correlation angle occurring around 0.2 fm$^{-1}$ is clearly above 90$^\circ$, even taking into account the errors.
This constitutes the first experimental evidence supporting universality of the dineutron correlation in the low-density nuclear surface of Borromean nuclei, which had been previously suggested \cite{mat06, hag07, kub20}. 
It is worth noting that, depending on the probe, an inclusive measurement of the correlation angle will be sensitive to a rather large region of the nucleus (including the interior), and the dineutron correlation signal may be damped, as shown in Fig.~\ref{f:theo-angle-dist}.

\section{\label{sec:level4}Theoretical analysis}
\begin{figure}[ht]
\begin{center}
\includegraphics[width=0.9\linewidth]{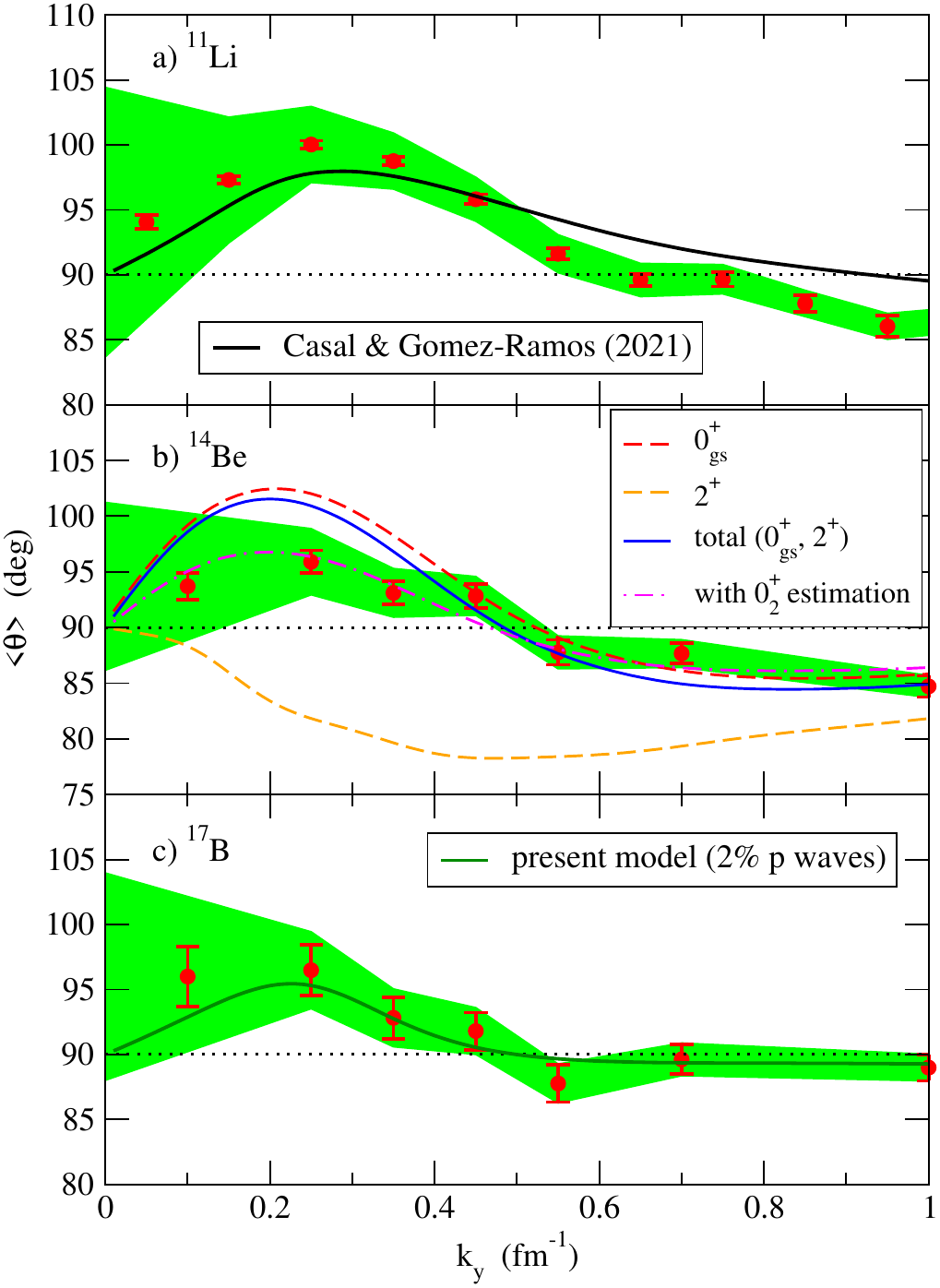}
    \caption{\label{f:theo-oa-vs-kmiss} Average correlation angle as a function of the intrinsic momentum for (a) $^{11}$Li \cite{cas21}, (b) $^{14}$Be and (c) $^{17}$B, compared to the theoretical calculations.}
 \end{center}
\end{figure}

To quantify and understand the mechanism behind the onset of the dineutron correlation, we compare the experimental result of Fig.~\ref{f:angle} to theoretical predictions in Fig.~\ref{f:theo-oa-vs-kmiss}. The theoretical description combines three-body models within the hyperspherical framework to describe the structure of Borromean nuclei \cite{JCasal13,Singh20} and an eikonal sudden description of the $(p,pN)$ reaction \cite{kik16}. This description has been used to study $^{11}$Li$(p,pn)$ in \cite{cas21}. We refer the reader to \cite{cas21} for a detailed overview of the model.
The structure model for $^{11}$Li was originally introduced in Ref.~\cite{gomezramosplb17} to describe GSI $(p,pn)$ data \cite{aks13a}, and later revisited in \cite{cas21} to incorporate $d$-wave contributions. 
This model leads to a large admixture between $s$ and $p$ waves ($\sim$ 60\% and 30\%, respectively), and the computed core-$nn$ rms distance is 4.9 fm, which compares well with the experimental value derived from Coulomb Dissociation data~\cite{nak06}.
For $^{14}$Be, we adopt the model in Ref.~\cite{cor19}, which is dominated by a low-lying $p$-wave resonance in $^{13}$Be ($\sim$ 60\% of the wave function comes from $p$ waves) and includes the effect of the first 2$^+$ excited state of the $^{12}$Be core (which amounts to roughly 20\% of the norm of the ground state of $^{14}$Be). For $^{17}$B, the three-body wave function was computed by fixing a simple model neglecting the spin of the core, in the same spirit as the $^{19}$B calculations in Ref.~\cite{JCasal20}, with the low-lying $s$ and $d$ states adjusted to reproduce the main features reported in the recent experimental work~\cite{yan21}. In such a model, the wave function is mostly governed by the $d_{5/2}$ component ($\sim$ 80\%), and the $p$-wave admixture is minimal ($\lesssim 2\%$) and comes from the non-resonant continuum in $^{16}$B. 
The calculated matter radii for $^{14}$Be and $^{17}$B are 3.0 and 2.8 fm, respectively, which compare well with the values reported in Ref.~\cite{suz99} from interaction cross sections.

Using these structure inputs, the calculations capture the general trend of the average correlation angle as a function of the intrinsic momentum, as shown in Fig.~\ref{f:theo-oa-vs-kmiss}.
Figure \ref{f:theo-oa-vs-kmiss}a corresponds to $^{11}$Li and was already explored in \cite{cas21} with the same theoretical description. It should be remarked that for missing momenta $k_y \gtrsim 0.5$ fm$^{-1}$, the distribution is affected by the core-proton interaction, so it is unreliable to extract nuclear structure information from that region \cite{cas21}.
In the case of $^{14}$Be (Fig. \ref{f:theo-oa-vs-kmiss}b), the calculated average correlation angle (blue solid line) follows the trend of the experimental data but the results for intrinsic momenta smaller than 0.5 fm$^{-1}$ are somewhat overestimated. 
Meanwhile, for $^{17}$B (Fig.~\ref{f:theo-oa-vs-kmiss}c), the theoretical model 
describes the maximum even with only a 2\% $p$-wave admixture. This remarkable sensitivity of the maximum to small opposite-parity components was already noted in \cite{cas21}. 

Only for $^{14}$Be there are significant differences between theoretical calculation and experimental data. To understand these differences, we note that in the analysis of the $^{13}$Be energy distribution in \cite{cor19} the three-body model used in this work was suggested to be missing some core-excited components. Different components of the $^{12}$Be core can give opposing contributions to the average correlation angle, as shown in Fig.~\ref{f:theo-oa-vs-kmiss}b, where the $^{12}$Be$(0^+_{gs})$ component's distribution (red dashed) goes over 90$^\circ$ at low momenta, while the excited $^{12}$Be$(2^+)$'s contribution (orange dashed) goes under 90$^\circ$. Among the missing components in the used model, those where the $^{12}$Be core is in its first excited $0^+_2$ state are particularly significant, since they are more likely to be populated, as its angular momentum and parity are those of the $^{14}$Be ground state. To estimate their effect, we note that the $^{12}$Be$(0^+_2)$ state is usually described as an orthogonal partner of the $0^+$ ground state \cite{mac18,for12}, with opposite relative sign between its positive and negative-parity components when compared to $^{12}$Be$(0^+_{gs})$. Therefore the components with $^{12}$Be($0^+_2$) should present a correlation angle smaller than 90$^\circ$ (opposite to $^{12}$Be$(0^+_{gs})$). Tentatively, for the correlation angle as a function of missing momentum, we have assigned to the $^{12}$Be($0^+_2$) components a distribution equal to that of $^{12}$Be$(0^+_{gs})$ but mirrored around 90$^\circ$, and a weight of 16\%, similar to the 20\% obtained with the three-body model for the similar-energy $^{12}$Be$(2^+)$. This estimation produces the magenta dot-dashed line, whose agreement with the data is much improved, pointing to the excitation of the core having a significant effect in the dineutron correlation, which was already indicated in \cite{kob16}. 
Therefore, the effect of the core may be responsible for $^{14}$Be and $^{17}$B showing a similar correlation angle, despite their very distinct admixture of different-parity components.


At this point, a natural question arises about how to compare the degree of dineutron correlation among different nuclei. A possible criterium is based solely on the experimental results, by comparing the maximum correlation angle. The maximum correlation angle for $^{11}$Li, $^{14}$Be and $^{17}$B occurs at $k_y=0.25$ fm$^{-1}$ and corresponds to 100.0(2)$^{+29}_{-29}$, 95.9(10)$^{+29}_{-29}$ and 96.4(19)$^{+29}_{-29}$ degrees, respectively. 

\begin{figure*}[ht]
\begin{center}
\includegraphics[width=0.33\linewidth]{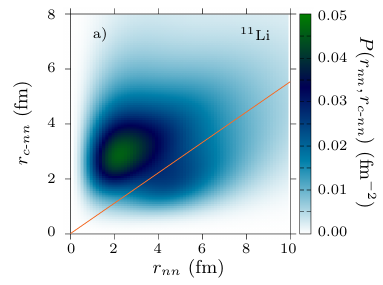}
\includegraphics[width=0.33\linewidth]{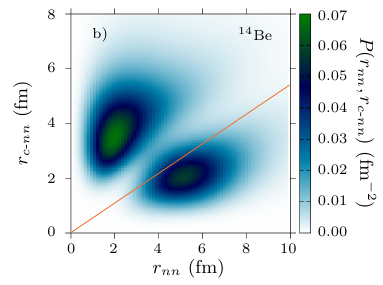}
\includegraphics[width=0.33\linewidth]{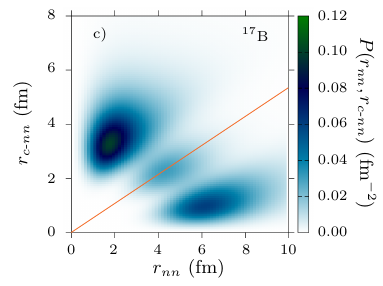}
    \caption{\label{f:theo-xy-density}Ground-state probability density, as a function of the distance between the halo neutrons ($r_{nn}$) and that between the center of mass of the $nn$ pair and the core ($r_{c\text{-}nn}$) for a) $^{11}$Li, b) $^{14}$Be and c) $^{17}$B. In a purely non-correlated case, the probability would be symmetric around the orange lines.}
 \end{center}
\end{figure*}

A second criterion is to make use of the theoretical models employed. 
The theoretical calculations in Fig.~\ref{f:theo-oa-vs-kmiss} give the average maximum values of 98.0 ($^{11}$Li), 96.6 ($^{14}$Be) and 95.4 ($^{17}$B) degrees, which compare well to the experimental results. It is worth noting that the corresponding three-body models give rise to maximum of the two-neutrons wave function density around the minimum of the average interneutron distance, as discussed in Ref.~\cite{hag07}, and this feature is directly linked to the present observations in momentum space.
The three-body model allows also to draw the ground-state probability density as a function of the Jacobi-$T$ coordinates r$_{nn}$ and r$_{c-nn}$, i.e. the distance among the two neutrons and the two neutrons baricenter with respect to the core. This is shown in Fig. \ref{f:theo-xy-density} for the three cases considered and allows to gain more insight on the configuration of the neutrons. In a purely non-correlated scenario, the distributions would present equal weights at both sides of the orange lines in the figure, which delimit two distinct regions within the hyperspherical description of three-body nuclei~\cite{Zhukov93}. Local maxima above this line, i.e., for small $r_{nn}$, are usually associated to the dineutron configuration, whereas the peaks below it correspond to the so-called ``cigar''-like structure. The dominance of one of these structures is associated to correlations. We can see that a clear dineutron peak is obtained for the three nuclei. For $^{11}$Li (Fig. \ref{f:theo-xy-density}a) the dineutron peak is clearly dominant, with only a relatively small fraction of the probability exploring larger neutron-neutron distances. For $^{14}$Be (Fig. \ref{f:theo-xy-density}b), the two configurations are clearly separated, with the dineutron still being more pronounced. In the case of $^{17}$B (Fig. \ref{f:theo-xy-density}c), three maxima appear (this is a consequence of the dominant $d$-wave content of the ground state). 

To quantify the degree of dineutron development for each nucleus, we may define the quantity
\begin{equation}
    \chi=\frac{P_d-P_c}{P_d+P_c},
    \label{eq:chi_dineutron}
\end{equation}
where $P_d$ and $P_c$ are the integrated probabilities above and below the symmetry lines in Fig. \ref{f:theo-xy-density}, i.e., $P_d$ is somehow a measure of the dineutron component, while $P_c$ is related to the cigar component. Indeed, with this definition $\chi=1$ ($-1$) would correspond to a ``pure'' dineutron (cigar). The integration for $^{11}$Li, $^{14}$Be and $^{17}$B within the present calculations yields $\chi=0.43$, $0.32$ and $0.19$, respectively. In this case, both criteria agree and support the fact that the dineutron correlation is stronger for $^{11}$Li.  
The theoretical model also permits the extraction of the average opening angle in configuration space, obtaining $\left\langle\theta_r\right\rangle=66.9^\circ (^{11}\mathrm{Li})$, $67.1^\circ (^{14}\mathrm{Be})$ and $77.4^\circ (^{17}\mathrm{B})$. The results for $^{11}\mathrm{Li}$ and $^{14}\mathrm{Be}$ are consistent to those presented in \cite{ber07}, while the angle for $^{17}\mathrm{B}$ is similar to that presented for $^{6}\mathrm{He}$.  Since both nuclei have very little admixture of different-parity components, their opening angles should be comparable. 

%

\begin{figure}[ht]
\begin{center}
\includegraphics[width=0.9\linewidth]{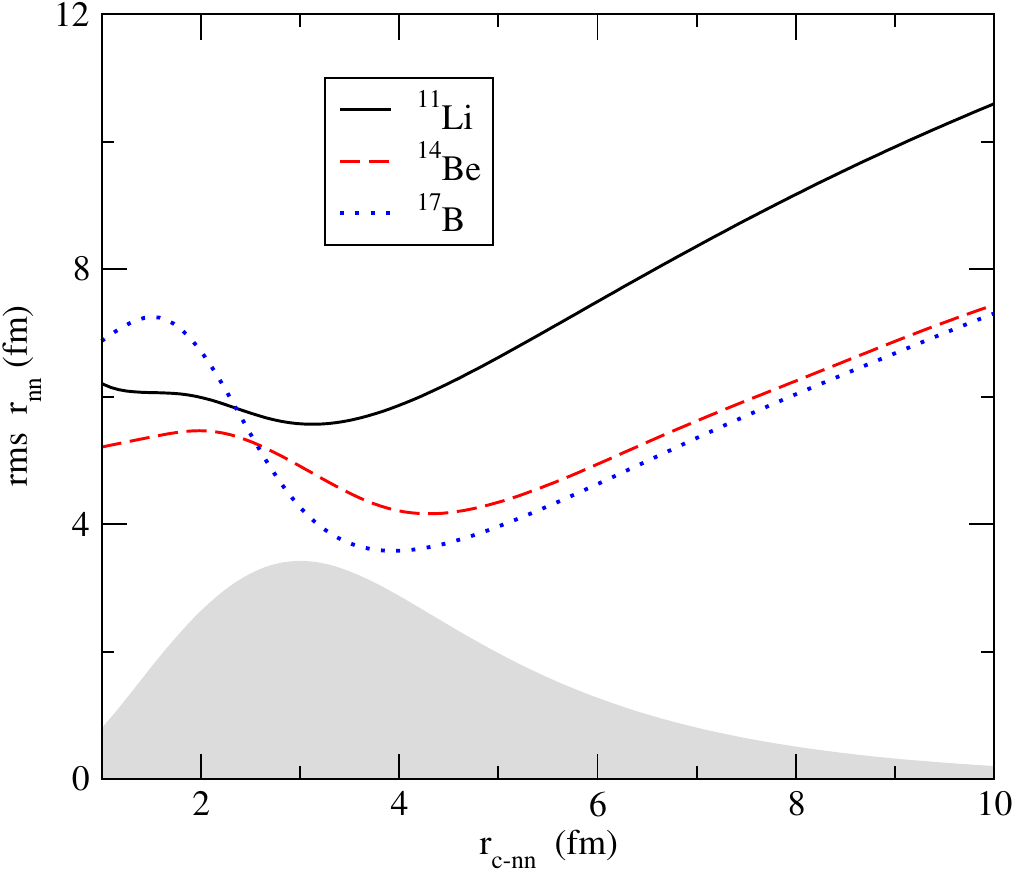}
     
    \caption{\label{f:theo-rxmin} Root mean square inter-neutron distance r$_{nn}$ as a function of the distance between the core and the center of mass of the two neutrons r$_{c-nn}$ for $^{11}$Li, $^{14}$Be and $^{17}$B. The shaded area represents the probability distribution of the two neutrons in $^{11}$Li, obtained by direct integration in Fig. \ref{f:theo-xy-density}.}
 \end{center}
\end{figure}

From Fig. \ref{f:theo-rxmin}, one can extract the correlation between the average r$_{nn}$ and r$_{c-nn}$. The minimum of r$_{nn}$ corresponds to a dineutron configuration, and its position signals the region of the nucleus where the calculation predicts the dineutron correlation to be stronger. We can notice that this occurs for r$_{c-nn}=3\text{-}4$ fm, corresponding to the nuclear periphery, again supporting the results in \cite{kub20} and generalizing them to $^{14}$Be and $^{17}$B. As discussed in~\cite{hag07}, this behaviour can be interpreted as a transition from BCS-like correlations in the interior to a BEC-like one, the spatially compact dineutron, around the surface. The probability density for the two valence nucleons of $^{11}$Li is also displayed with a grey area. One can notice that the probability maximum and the inter-nucleon distance minimum are attained around the same value of r$_{c-nn} \sim 3$ fm, which makes the dineutron configuration dominant for the two valence nucleons. Within the adopted theoretical framework, the maximum in r$_{c-nn}$ corresponding to the nuclear surface can be associated to the maximum at low intrinsic momentum k$_y$, validating the use of k$_y$ as a proxy for peripherality.

\section{\label{sec:level5} Conclusions}
 We have presented a comparative study of dineutron correlation in three Borromean systems, $^{11}$Li, $^{14}$Be and $^{17}$B, based on the average correlation angle as a function of the intrinsic momentum of the removed neutron. This work follows the seminal work of Kubota \emph{et al.} \cite{kub20} who first proposed to use this observable to probe the location of dineutron correlation inside the nucleus, and extends the study to $^{14}$Be and $^{17}$B. A dineutron correlation appears in the periphery of $^{14}$Be and $^{17}$B as well, but is damped compared to $^{11}$Li. \\ This study provides the first experimental hint of the universality of dineutron correlation in the low-density surface of Borromean nuclei. Even while fast nucleon removal induced by high-energy quasi-free scattering is the tool of choice to reduce the effect of final-state interactions, consistent measurements using different probes may help to confirm the universal character of our observation. The damping of dineutron correlation in $^{14}$Be is interpreted as due to the presence of configurations with an excited core, that can be predicted within the three-body model. Higher statistics data incorporating gamma-ray coincidences, which enable core
excitations to be probed, could be used to investigate this explanation.


\section*{Acknowledgements}
This work has been supported by the European Research Council through the ERC Starting Grant No. MINOS-258567. 
J.C., M.G.R.\ and A.M.M.\ acknowledge financial support by MCIN/AEI/10.13039/501100011033 under I+D+i project No.\ PID2020-114687GB-I00 and under grant IJC2020-043878-I (also funded by ``European Union NextGenerationEU/PRTR''), by the European Union's Horizon 2020 research and innovation programme under the Marie Skłodowska-Curie grant agreement No.\ 101023609, by the Consejer\'{\i}a de Econom\'{\i}a, Conocimiento, Empresas y Universidad, Junta de Andaluc\'{\i}a (Spain) and ``ERDF-A Way of Making Europe'' under PAIDI 2020 project No.\ P20\_01247, and by the European Social Fund and Junta de Andalucía (PAIDI 2020) under grant number DOC-01006.  J.G., F.M.M. and N.A.O. acknowledge partial support from the Franco-Japanese LIA-International Associated Laboratory for Nuclear Structure Problems as well as the French ANR14-CE33-0022-02 EXPAND. Z.K. and L.S. acknowledge partial support by the Institute for Basic Science (IBS-R031-D1). S.P. acknowledges the support of the UK STFC under contract numbers ST/L005727/1 and ST/P003885/1 and the Deutsche Forschungsgemeinschaft (DFG, German Research Foundation) Project-ID 279384907 - SFB 1245.

\bibliographystyle{h-physrev}
\bibliography{biblio2}

\end{document}